\documentclass{aa}

\usepackage{natbib}
\usepackage{longtable}
\usepackage{lscape}
\usepackage{amsmath}
\usepackage{txfonts}
\usepackage{color}
\usepackage{url}
\usepackage{xspace}

\usepackage{graphicx}
\bibpunct{(}{)}{;}{a}{}{,}

\newcommand{\kev}{keV\xspace}
\newcommand{\xmm}{\textsl{XMM-Newton}\xspace}
\newcommand{\rosat}{\textsl{ROSAT}\xspace}
\newcommand{\chandra}{\textsl{Chandra}\xspace}

\begin{document}
   \title{A Wolf-Rayet/black-hole X-ray binary candidate in NGC~300}

   \author{S. Carpano
          \inst{1}
          \and
	  A.M.T.~Pollock
          \inst{1}
          \and
          J. Wilms
          \inst{2}
	  \and
	  M. Ehle
	  \inst{1}
	  \and
	  M. Schirmer
	  \inst{3}}

   \offprints{S. Carpano, e-mail: scarpano@sciops.esa.int}
   \institute{XMM-Newton Science Operations Centre, ESAC, European
   Space Agency, Apartado 50727, 28080 Madrid, Spain 
             \and
   Dr.\ Remeis Sternwarte, Astronomisches Institut der FAU Erlangen-N\"urnberg,
   Sternwartstr. 7, 96049 Bamberg, Germany 
             \and
   Isaac Newton Group of Telescopes, 38700 Santa Cruz de La Palma, Spain
   }
   
   \date{Submitted: 9 October 2006; Accepted: 6 November 2006}
   
   \abstract{Wolf Rayet/black hole binaries are believed to exist as a
     later evolutionary product of high-mass X-ray binaries. Hundreds
     of such binaries may exist in galaxies, but only a few of them are
     close enough to be observed as X-ray binaries. Only a couple of
     candidates have been reported so far.}{ Based on \xmm
     observations, we report the positional coincidence of the
     brightest X-ray source in NGC~300 (NGC~300~X-1) with a Wolf-Rayet
     candidate. Temporal and spectral analysis of the X-ray source is
     performed.}{We determine an accurate X-ray position of the
     object, and derive light curves, spectra and flux in four \xmm
     observations.}{ The positions of the X-ray source and the helium
     star candidate coincide within $0\farcs11\pm0\farcs45$. The X-ray
     light curves show irregular variability. During one \xmm
     observation, the flux increased by about a factor of ten in 10
     hours. The spectrum can be modelled by a power-law with
     $\Gamma\sim2.45$ with additional relatively weak line emission,
     notably around 0.95\,\kev. The mean observed (absorbed) luminosity in the
     0.2--10\,\kev band is
     $\sim2\times10^{38}\,\text{erg}\,\text{s}^{-1}$. }{ NGC~300 X-1 is
     a good candidate for a Wolf-Rayet/black-hole X-ray binary: its
     position coincides with a Wolf-Rayet candidate and the unabsorbed
     X-ray luminosity reached
     L$_{0.2-10\,\text{\kev}}\sim1\times10^{39}\,\text{erg}\,\text{s}^{-1}$,
     suggesting the presence of a black hole.}

     \keywords{Galaxies: individual: NGC~300 -- X-rays: binaries  -- Stars: Wolf-Rayet} 

\maketitle
%

\section{Introduction}
\label{sec:int}
A helium-star/compact-object X-ray binary was suggested by
\cite{vandenHeuvel1973} and \cite{vandenHeuvel1983} to explain the
nature of \object{Cyg X-3}, and discussed further by
\cite{vanKerkwijk1995}. Such systems are believed to be a late
evolutionary product of high-mass X-ray binaries. HMXBs are composed
of an early-type star and a neutron-star or black-hole (BH) compact object.
Material from the strong wind of the star is accreted onto the compact
object with the emission of X-ray radiation. When the secondary star
exhausts its core-hydrogen, it expands and begins to overflow its
Roche-Lobe. A common envelope forms and the compact object starts to
spiral in. If the hydrogen envelope is expelled before the compact
object has reached the helium core, a stable binary system with a
short orbital period is formed. \cite{Ergma1998} showed, by means of
population synthesis, that a few hundred such black holes with helium
star companions may form in our galaxy. The orbital periods of the
majority exceed 10\,hrs, with a maximum at $\sim$100\,hrs. However,
following \cite{Illarionov1975}, the formation of accretion disks from
the strong stellar wind of a Wolf-Rayet (WR) star is possible only for
short orbital periods of ($<$10\,hrs). The number of such systems is
expected to be extremely small \citep{Ergma1998}.

In the Galaxy, Cyg~X-3 is a good candidate for such a binary system.
This X-ray source is one of the most luminous in the Galaxy with
$L_\text{X}\sim10^{38}$\,erg~s$^{-1}$ and has a Wolf-Rayet companion
star designated WR~145a \citep{vanKerkwijk1992}.  A second candidate,
IC~10~X-1 ($L_\text{X}\sim1.2\times10^{38}$\,erg~s$^{-1}$) has been
observed in the starburst galaxy \object{IC~10}
\citep{Bauer2004,Wang2005}.  Although two other much less luminous
candidates in the LMC were suggested by \citet{Wang1995} from \rosat
observations, \cite{Portegies2002} and \citet{Townsley2006} concluded,
partially on the basis of more recent \chandra data, that these two
objects are more likely to be colliding-wind binaries.  In this paper
we report the discovery of a third high-luminosity candidate for a WR/BH
X-ray binary in the galaxy \object{NGC~300}.

NGC~300 is a dwarf spiral galaxy, belonging to the Sculptor galaxy
group, located at a distance of $\sim$1.88\,Mpc \citep{Gieren2005}. Because it is
almost face-on oriented and has a low Galactic column density of
$N_\text{H}=3.6\times10^{20}\,\text{cm}^{-2}$ \citep{Dickey1990}, it
is well suited for studies of stellar content. Based on deep VLT-FORS
narrow-band imaging, \cite{Schild2003} detected 58 WR star
candidates in the central region of the galaxy, of which 16 were
already spectroscopically confirmed \citep{Schild1991,Breysacher1997}.
Their technique consisted of taking images through two interference
filters with central wavelengths at $4684\AA$, containing WR
emission lines, and at $4781\AA$ as a continuum reference, with band
widths of $66\AA$ and $68\AA$, respectively. WR candidates were
selected as having peak intensities of at least $6\sigma$ in the
difference ($4684\AA$--$4781\AA$) image.

The X-ray source population of NGC~300 has been studied with \rosat
\citep{Read2001} and \xmm \citep{Carpano2005}. With mean observed
luminosities of
$L_{0.1-2.4\,\text{\kev}}\sim2.2\times10^{38}\,\text{erg}\,\text{s}^{-1}$
\citep{Read2001} and
$L_{0.3-6.0\,\text{\kev}}\sim1.7\times10^{38}\,\text{erg}\,\text{s}^{-1}$
\citep{Carpano2005}, the brightest  X-ray source has been
suggested to be a black hole of about $5M_\odot$.  In this Letter, we
report the positional coincidence of this source with one of the
WR candidates reported by \cite{Schild2003}. We refer to the
X-ray source as \object{NGC~300 X-1} and to the Wolf-Rayet star as WR-41
following \cite{Schild2003}. The remainder of the Letter is
organised as follows. Section~\ref{sec:obs} briefly describes the \xmm
observations and data reduction. In Sect.~\ref{sec:image}, we report
the spatial coincidence of NGC~300~X-1 and the WR~41. Temporal and
spectral analysis of the \xmm X-ray source is shown in
Sect.~\ref{sec:time_spec}, while a discussion of our results is given in
Sect.~\ref{sec:conc}.


\section{Observations and data reduction}
\label{sec:obs}

To date, NGC~300 has been observed  four times with \xmm
on 2000 December 26 for 32\,ksec EPIC-pn time;
on 2001 January 01 for 40\,ksec;
on 2005 May 22 for 35\,ksec; and
on 2005 November 25 for 35\,ksec
during revolutions 0192, 0195, 0998 and 1092, respectively.
For each observation, the EPIC-MOS
\citep{Turner} and EPIC-pn \citep{Strueder} cameras were operated in
full frame mode with the medium filter.  The data reduction
was identical to that performed in our analysis of the previous \xmm
observations \citep{Carpano2005,Carpano2006}, except that version 7.0.0 of the
\xmm Science Analysis System (SAS) and the most recent calibration files were used.

High fluxes of proton flares were observed during revolutions 0192 and
0998.  After screening the MOS data for flares using the standard
procedures described by the \xmm
team\footnote{\url{http://xmm.esac.esa.int/external/xmm_user_support/documentation/sas_usg/USG/}},
a total of 30\,ksec of low background data remained in each
revolution.  A more detailed description of the \xmm data reduction
can be found in \cite{Carpano2005} for the first two \xmm observations
and \cite{Carpano2006} for the last two.


\section{Position coincidence of NGC~300~X-1 and WR-41}
\label{sec:image}
An accurate \xmm X-ray position of NGC~300~X-1 was derived considering
both statistical and systematic errors.  Statistical errors depend on
the brightness of the source. For a source as bright as NGC~300~X-1,
they are small compared with the systematic errors in the X-ray
reference frames of each observation.  Although merging all four \xmm
observations would in principle reduce the statistical errors in the
absence of systematic errors, we chose to use only the data of
revolution 0195, where the source was bright and well separated from
local instrumental features, such as chip gaps, which affected the other 
observations to some extent.

Initial estimates of the position of NGC~300~X-1 and its statistical
error were derived using the SAS \texttt{edetect\_chain} task,
which performs maximum-likelihood source detection. The source was
detected with a likelihood $L$ of $44833$. Probabilities $P$, are 
related to maximum likelihood values $L$, by the relation $P= 1- \exp{(-L)}$.
The statistical position error is of $0\farcs083$ (or $0\farcs077$ with 
the merged data). The
systematic errors were tackled with the SAS task \texttt{eposcorr},
which correlates X-ray source positions with those of their optical
counterparts, as explained by \cite{Carpano2005}. For this, we used 12 sources
inside the galaxy disk that have clear optical counterparts.
The systematic shift
(in the sense X-ray minus optical position) was $-1\farcs{}25\pm 0\farcs{}28$
in right ascension and $-0\farcs{}35\pm 0\farcs{}34$ in declination.
The revised coordinates of NGC~300~X-1 are therefore
$\alpha_\text{J2000}=00^\text{h}55^\text{m} 10\fs{}00$ and
$\delta_\text{J2000}=-37^\circ 42' 12\farcs 06$, with an uncertainty
of $0\farcs 45$.

\begin{figure}
  \resizebox{\hsize}{!}{\includegraphics{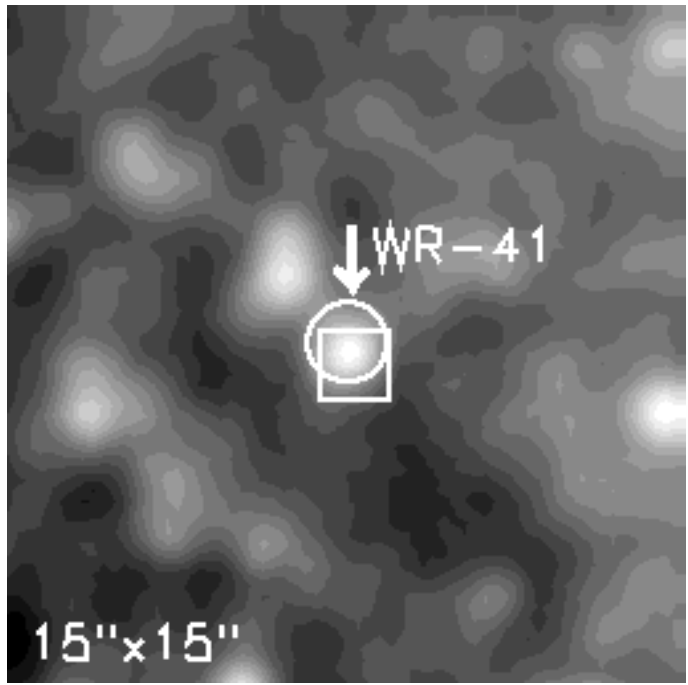}}
 \caption{$15''\times15''$ optical (B-band) image centred on the X-ray
   position of NGC~300~X-1. The circle represents the 2$\sigma$
   uncertainty of the X-ray position, while the box represents the
   position of WR-41 determined by \cite{Schild2003}. North is up 
   and East is left.}
 \label{fig:image}
\end{figure}

Figure~\ref{fig:image} shows the optical (B-band) image of the field
taken with the 2.2\,m MPG/ESO telescope on La Silla.  See
\citet{Carpano2005} and \citet{Schirmer2003} for a description of the
optical data and their reduction. The circle represents the 2$\sigma$
uncertainty of the X-ray position, while the box is centred on the
position of WR-41, derived by \cite{Schild2003}. We redetermined the
coordinates of the Wolf-Rayet star in the broad B-band image, using the IDL \texttt{find}
routine, which is part of the \texttt{idlphot} photometry
library\footnote{\url{http://idlastro.gsfc.nasa.gov/contents.html}}
implemented from an early Fortran version of the DAOPHOT aperture
photometry package \citep{Stetson1987}.  The revised coordinates of
WR-41 are $\alpha_\text{J2000}=00^\text{h}55^\text{m} 09\fs{}99$ and
$\delta_\text{J2000}=-37^\circ 42' 12\farcs 16$.  The absolute astrometric
accuracy of the optical image is about $0\farcs 25$.
The difference between the position of NGC~300~X-1
and the revised position of WR-41 is thus $0\farcs11\pm0\farcs51$ (or $0\farcs52$
using the position of \cite{Schild2003}). The spatial coincidence is
comparable to that found for \object{IC~10~X-1},
\citep[$0\farcs23\pm0\farcs30$,][]{Bauer2004}, which lies  at a distance of
0.8\,Mpc.


\section{Timing and spectral analysis of NGC~300~X-1}
\label{sec:time_spec}

\begin{figure}
  \resizebox{\hsize}{!}{\includegraphics[bb=20 26 333 486,clip=true]{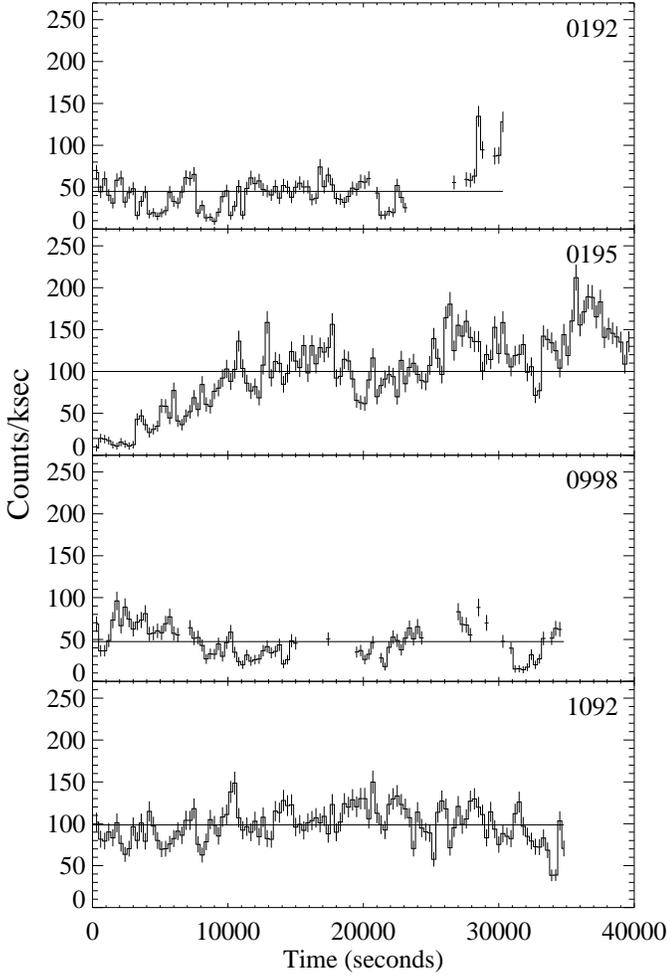}}
 \caption{EPIC-pn and -MOS 0.2-10.0\,\kev light curve of 
   NGC~300~X-1 in revolutions 0192, 0195, 0998 and 1092 from top to
   bottom.  Periods of high background have been excluded from the
   data.  The horizontal lines show the mean values. 
   Times start from the beginning of each observation.}
 \label{fig:light}
\end{figure}

Figure~\ref{fig:light} shows the \xmm light curves of NGC~300~X-1
separately for revolutions 0192, 0195, 0998 and 1092, incorporating
background-corrected MOS and pn data.  The gaps in the data of
revolutions 0192 and 0998 correspond to periods of high  soft-energy
proton flux.  The time bin size is 300 sec. The light curve of
NGC~300~X-1 is modulated by short-term (few 1000 sec), large-amplitude
(factors of $\sim$4-5) variations in all four observations. The same
type of short- and long-term variations has been reported for IC~10~X-1
\citep{Bauer2004,Wang2005}. During revolution 0195, the flux increased
by a factor of 10 within 10\,hrs, and the highest observed luminosity, 
assuming isotropic emission, was
$L_{0.2-10.0\,\text{\kev}}\sim6\times10^{38}\,\text{erg}\,\text{s}^{-1}$.  
The corresponding unabsorbed luminosity
is $\sim1\times10^{39}\,\text{erg}\,\text{s}^{-1}$. Using periodograms
and epoch-folding, no short-time periodic signal was found on
timescales between 5\,sec and 30\,ksec.

\begin{figure}
  \resizebox{\hsize}{!}{\includegraphics[angle=-90]{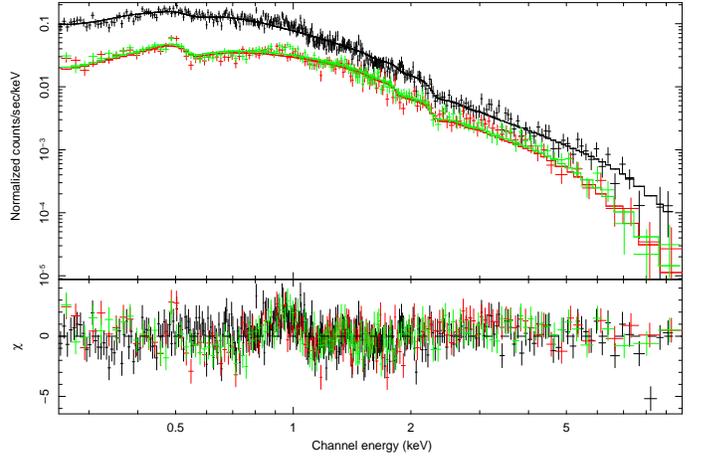}}
 \caption{EPIC-pn (black) and -MOS (red and green) spectra of NGC~300~X-1 added for the four \xmm observations. The spectrum
   is fitted with an absorbed power-law model. Bottom: residuals expressed in $\sigma$. Note the excess in the residuals
   around 0.95\,\kev suggesting the presence of discrete emission.}
 \label{fig:spec}
\end{figure}

Figure~\ref{fig:spec} shows the pn and MOS spectra of NGC~300~X-1. For
clarity, we have plotted the total spectra from the combination of all
four observations. We first tried to fit single power-law models to
the data, where parameters were left free in each observation, yielding
$\chi^2_{\nu}/\nu=1.32$.  From the residuals shown at the bottom of
Fig.~\ref{fig:spec}, some excess around 0.95\,\kev is evident. Adding
a Gaussian line significantly improved the fit,
$\chi^2_{\nu}/\nu=1.15$ for $\nu=1033$ degrees of freedom. The
best-fitting parameters of the absorbed combined power-law and single
Gaussian-line model are shown in Table~\ref{tab:spec_fit} for each
observation separately. $N_\text{H}$ is the
equivalent column density of neutral hydrogen, $\Gamma$ the photon
index, $E_{\text{L}}$ the energy of the line and
$\sigma_{\text{L}}$ its width, and $\text{Norm}_{\text{L}}$ and
$\text{Norm}_{\Gamma}$ are the normalisation constants for the
Gaussian line and the power-law component, respectively. The
corresponding 0.2--10\,\kev flux, absorbed and
unabsorbed luminosities are shown in the last three
rows. Uncertainties are given at 90\% confidence level.
Allowing a variable photon index between
observations does not improve the fit or change spectral parameters
within the errors except for revolution 0998, where 
$N_\text{H}=4.78^{+1.06}_{-1.36}\times10^{20}\,\text{cm}^{-2}$ and
$\Gamma=2.35^{+0.08}_{-0.09}$.

The mean observed luminosity increased by more than a factor of two
from the lowest state in revolution 0192 to the highest in revolution
1092. When the source was brighter, the intrinsic absorption, as well
as the emission line/continuum ratio, increased. In revolution 0192,
the equivalent hydrogen column, $N_\text{H}$, was compatible with the
galactic value ($N_\text{H}=3.6\times10^{20}\,\text{cm}^{-2}$),
implying little or no intrinsic absorption in the low state.

We also tried to fit the data using a power-law model combined with thermal
emission from a collisionally ionized plasma modelled by \texttt{apec}
in XSPEC (version 11.3.2). This thermal emission includes both lines and continuum
emissivities with a best-fit temperature of $\sim$0.86$^{+0.03}_{-0.04}$\,\kev. This
model is statistically indistinguishable from the power-law plus
Gaussian line model.

\begin{table}
 \centering
 \caption{Results of the spectral fits for NGC~300~X-1, using an
   absorbed power-law model and a Gaussian line
   (\texttt{phabs*(gauss+power)}, in XSPEC). } 
  \label{tab:spec_fit}
 \begin{tabular}{lllll}
 \hline  
 \xmm rev.: & 0192 & 0195 & 0998 & 1092\\
 \hline
 $N_\text{H}(\times 10^{20}\,\text{cm}^{-2})$& 4.22$^{+0.61}_{-0.71}$& 7.84$^{+0.57}_{-0.69}$& 6.04$^{+0.79}_{-0.44}$& 7.79$^{+0.60}_{-0.69}$\\[3pt]
 $\Gamma$& \multicolumn{4}{c}{$\,2.45^{+0.03}_{-0.03}\,$}\\[3pt]
 Norm$_{\,\Gamma}\times 10^{-4}$ & 0.55$^{+0.02}_{-0.02}$& 1.43$^{+0.04}_{-0.05}$& 1.21$^{+0.05}_{-0.06}$&1.45$^{+0.05}_{-0.05}$\\[3pt]
 E$_{\text{\,L}}$ (\kev)& \multicolumn{4}{c}{$\,0.94^{+0.01}_{-0.01}\,$}\\[3pt]
 $\sigma_{\text{\,L}}$ (\kev)& \multicolumn{4}{c}{$\,0.07^{+0.01}_{-0.01}\,$}\\[3pt]
 Norm$_{\text{\,L}}\times 10^{-5}$ &$<0.10$ & 1.13$^{+0.35}_{-0.18}$& 1.00$^{+0.39}_{-0.32}$& 1.60$^{+0.30}_{-0.29}$\\[3pt]
 $F_{0.2-10\,\text{\kev}}\times 10^{-13}$ (\,cgs) & 2.36$^{+0.14}_{-0.16}$ & 5.40$^{+0.19}_{-0.21}$& 4.87$^{+0.29}_{-0.30}$& 5.56$^{+0.23}_{-0.23}$\\[3pt]
 $L_{0.2-10\,\text{\kev}}^{\text{obs}}\times 10^{38}$ (\,cgs) &1.00$^{+0.06}_{-0.07}$ &2.29$^{+0.08}_{-0.09}$ &2.06$^{+0.12}_{-0.13}$& 2.35$^{+0.10}_{-0.10}$\\[3pt]
 $L_{0.2-10\,\text{\kev}}^{\text{unabs}}\times 10^{38}$ (\,cgs) & 1.43$^{+0.05}_{-0.12}$ & 3.75$^{+0.06}_{-0.21}$& 3.18$^{+0.11}_{-0.26}$& 3.85$^{+0.08}_{-0.22}$\\[3pt]
 \hline
   \end{tabular}
\end{table}


\section{Discussion}
\label{sec:conc}

WR-41 is a Wolf-Rayet candidate discovered by \cite{Schild2003}. Using
narrow band imaging techniques, they measured that
the flux of the source at $\lambda=4684\pm66\AA$, where strong WR
emission lines are observed, has an excess of 0.95\,mag over the
continuum. Its relative magnitude is
$m_\text{V}=22.44\,\text{mag}$, which corresponds, at a distance of
1.88\,Mpc, to an absolute magnitude of $-3.93\,\text{mag}$. 
Following \cite{Schild2003}, objects such as WR-41 are
candidates for weak-line single WN-type stars.

We have shown in this Letter that the brightest X-ray source in\
the nearby galaxy NGC~300, NGC~300~X-1, is likely to be the X-ray counterpart of
WR-41. This would imply that the system is a good candidate for
membership in the rare class of WR/BH X-ray binaries. Irregular flux
variations were observed in the four light curves, with a particularly
large increase observed over about 10 hours during \xmm revolution
0195. The origin of these variations is not clear, but likely due to
variations in the accretion rate and/or variable absorption in the
strong stellar wind of the donor star.
We also note that no obvious eclipse is visible in any of the light
curves.  As WR/BH X-ray binaries should have small orbital periods
$<$10\,hrs, we interpret the absence of eclipses as a possible
indication of face-on orientation of the system.

In the context of a WR/BH X-ray binary, the weak discrete emission
observed in the spectra of NGC~300-X-1, which is likely to be composed
of several unresolved lines, may arise from reprocessing by the
photoionized stellar wind: the X-rays emitted around the black hole
are ionizing the high density wind of the Wolf-Rayet star. This may
explain why this emission is more pronounced when the X-ray flux is higher.
The dense wind is probably also responsible for the intrinsic
absorption that increases as the X-ray source gets brighter.  Similar
ideas explain the weak emission lines observed in Cyg~X-3
\citep[e.g.,][]{Kawashima1996, Paerels2000} and in other wind accretors.
Faint lines have also been observed in the X-ray spectrum of the
candidate IC~10~X-1 \citep{Bauer2004}.  We note that the contribution
of the WR star to the X-ray flux is negligible as typical 
X-ray luminosities are about $10^{32}\,\text{erg}\,\text{s}^{-1}$ for single stars 
and $10^{34}\,\text{erg}\,\text{s}^{-1}$ for binaries
\citep[e.g.,][]{Pollock1987}.

To conclude, NGC~300~X-1 and WR-41 are good candidates for membership
in the very rare class of Wolf-Rayet/black hole X-ray binaries. This
conclusion is supported by the spatial coincidence of the sources as
well as the high maximum X-ray luminosity near
$10^{39}\,\text{erg}\,\text{s}^{-1}$.  As the nature of WR-41
currently relies on narrow-band photometric techniques, we encourage
optical observers to acquire the spectra necessary to confirm the
identification of WR-41 as a Wolf-Rayet star.

\begin{acknowledgements}
  This paper is based on observations obtained with \textsl{XMM-Newton}, an ESA
  science mission with instruments and contributions directly finded
  by ESA Member States and NASA, and on observations
  made with ESO Telescopes at the La Silla observatory and retrieved
  from the ESO archive. 
\end{acknowledgements}

\end{document}